\documentclass{iopart}

\jl{6}        
\eqnobysec    

\def\beq{\begin{equation}}
\def\eeq{\end{equation}}

\def\dualp#1{{}^{\ast_{(\hbox{$\scriptstyle #1$})}} \kern-1pt}
\def\dual{\,{}^\ast{}\kern-1.5pt}
\def\leftdual{\,{}^\ast{}\kern-1.5pt}
\def\rightdual{\leftdual}
\def\rightduals{\leftdual}
\def\dualp#1{{}^{\ast_{(\hbox{$\scriptstyle #1$})}} \kern-1pt}

\def\del{\nabla}
  \def\del{\nabla}
\def\div{\mathop{\rm div}\nolimits}    \def\curl{\mathop{\rm curl}\nolimits}
\def\Scurl{\mathop{\rm Scurl}\nolimits} 
\def\TF{{}^{(\rm TF)}}

\def\Lie{\hbox{\it\char'44}} \let\pounds=\Lie

\def\oversymbol#1#2{\vbox{\ialign{##\crcr \hfil$#1$\hfil\crcr
   \noalign{\kern1pt\nointerlineskip}%
   \hbox{$\hfil\displaystyle#2\hfil$}\crcr}}}
\def\overcirc#1{\oversymbol{\scriptstyle\kern.5pt \circ}{#1}}
\def\four{{}^{(4)}\kern-1pt}

\catcode`@=11     

\def\meqalign#1{\null\,\vcenter{\openup\jot\m@th
  \ialign{\strut\hfil$\displaystyle{##}$&&$\displaystyle{{}##}$\hfil
      \crcr#1\crcr}}\,}

\catcode`@=12   

\begin{document}

\title[The Cotton, Simon-Mars and Cotton-York Tensors in Stationary Spacetimes
]
{The Cotton, Simon-Mars and Cotton-York Tensors in Stationary Spacetimes}

\author{
Donato Bini${}^{\dag\,\ddag}$,
Robert T. Jantzen${}^{\S \,\ddag}$,
Giovanni Miniutti${}^{\P \,\ddag}$
}
\address{
  ${}^{\dag}$\
  Istituto per Applicazioni della Matematica C.N.R.,
  I--80131 Napoli, Italy
}
\address{
  ${}^{\ddag}$\
  International Center for Relativistic Astrophysics,
  University of Rome, I--00185 Roma, Italy
}
\address{
  ${}^{\S}$\
  Department of Mathematical Sciences, Villanova University,
  Villanova, PA 19085, USA
}
\address{
  ${}^{\P}$\
  Dipartimento di Fisica, University of Rome ``La Sapienza'', I--00185 Roma, Italy
}

\begin{abstract}
The Cotton-York and Simon-Mars tensors in stationary vacuum spacetimes are studied in the language of the
congruence approach pioneered by Hawking and Ellis.
Their relationships with the Papapetrou field defined by the stationary Killing congruence
and with a recent characterization of the Kerr spacetime in terms of the alignment between of the
principal null directions of the Weyl tensor with those of the  Papapetrou field are also investigated in this more transparent language.
\end{abstract}

\pacno{04.20.Cv}
\submitted July 19, 2001 \qquad 
Revised September 27, 2001

\section{Introduction}

The Kerr and Schwarzschild solutions have been characterized within the class of stationary asymptotically flat vacuum spacetimes
by the vanishing of the Simon tensor \cite{Simon}, recently reformulated as a spacetime tensor by Mars \cite{Mars99}.
In the Schwarzschild case,
the vanishing of the Simon tensor is equivalent to the vanishing of the
Cotton-York tensor \cite{York} associated with the conformally flat quotient geometry of the stationary Killing congruence.
In the Kerr case, the Cotton-York tensor is nonzero and the vanishing of the Simon tensor
has only recently been explained \cite{Mars00,Mars01,Fasop99, Fasop00} in terms of the alignment of the two principal
null directions associated with the Papapetrou field defined by the timelike Killing congruence and the two repeated
principal null directions of the Weyl tensor.

These questions center around the Bianchi identities for the Weyl tensor in vacuum, which involve the spacetime Cotton tensor. The congruence spacetime splitting approach pioneered by Hawking and Ellis \cite{HE,Ellis73} permits an elegant decomposition of these equations into Maxwell-like form for the electric and magnetic parts of the Weyl tensor \cite{Ellis99,Maartens}. This is the curvature level manifestation of gravitoelectromagnetism, the nonlinear analogy between general relativity and electromagnetism, while at the connection level Einstein's equations themselves based on metric potentials (lapse, shift, spatial metric) for the connection provide an underlying foundation \cite{Mol,Thorne,mfg,idcfii}. Exact solution theory for stationary spacetimes \cite{Kramer} developed its own peculiar notation and choice of variables adapted to exact solution techniques; these can be translated into the gravitoelectromagnetic language, clarifying the meaning of the quantities involved. This is helpful with the Simon-Mars discussion.

Starting with the Bianchi identities for the Weyl tensor in a vacuum stationary spacetime, conveniently written in their gravitoelectromagnetic form with respect to a congruence of Killing observers and expressed using the complex self-dual formalism, one immediately sees the importance of the Simon-Mars tensor and its role in aligning the geometry at the connection and curvature levels when it vanishes, and the relation between the spacetime Cotton tensor and the spatial Cotton-York tensor.
We obtain  a number of new characterizations of the
Kerr spacetime involving the Cotton-York tensor which are equivalent to the vanishing Simon-Mars tensor condition. The more transparent language of this approach also makes it easier to interpret the quantities which are central to the discussion.
Some useful definitions and notation are given in the appendices.

\section{$1+3$ splitting of the spacetime and gravitoelectromagnetism}

A spacetime which admits a congruence of timelike lines (world lines of the test observers) with unit tangent vector (4-velocity)
$u$ ($u_\alpha u^\alpha =-1$) can be split into space-plus-time through the
orthogonal decomposition of each tangent space into a local time direction
along $u$ and the orthogonal local rest space LRS$_u$.
Tensors and tensor fields with no components along $u$ are called spatial
with respect to $u$. The orthogonal decomposition representation of any tensor field is accomplished by the
application of the identity expressed in terms of the associated temporal and spatial projections to each index of the tensor
\begin{eqnarray}
 \delta^\alpha{}_{\beta} &=&  T(u)^\alpha{}_\beta + P(u)^\alpha {}_\beta
\ ,\nonumber\\ 
 T(u)^\alpha{}_\beta &=& - u^\alpha u_\beta \ ,\qquad
 P(u)^\alpha {}_\beta = \delta^\alpha{}_{\beta} + u^\alpha u_\beta
\label{P(u)}\ .
\end{eqnarray}
The measurement of a generic tensor is defined as the set of all spatial
tensors obtained through the projection operations.
The congruence $u$ is itself characterized by the kinematical quantities (acceleration vector, expansion tensor: trace expansion scalar plus tracefree shear tensor, vorticity tensor) 
which result from the splitting of its covariant derivative $\nabla u$
\begin{eqnarray}
 a (u)_\alpha &=& u_{\alpha\,;\,\beta}\,u^\beta \ ,\nonumber \\
 \theta (u)_{\alpha\beta} &=& P(u)^\gamma {}_\alpha\,
  P(u)^\delta {}_\beta\,u_{\,(\,\gamma\,;\,\delta\,)}\ ,\qquad \hbox{\rm Tr}\, \theta(u)= \Theta (u), \nonumber \\
 \omega (u)_{\alpha\beta} &=&- P(u)^\gamma {}_\alpha\,
  P(u)^\delta {}_\beta\,u_{\,[\,\gamma\,;\,\delta\,]} \ .
\end{eqnarray}

Stationary spatial tensor fields correspond in a natural way to tensor fields on the Riemannian manifold of Killing observer world lines, the quotient space of the spacetime by the 1-parameter group of diffeomorphisms of the Killing vector field $u$, where it is timelike. This isomorphism relates the Mars spacetime Simon tensor (defined independently of the causal nature of the stationary Killing vector) to the original Simon tensor defined on the quotient space for Killing observers (which in contrast exists only where the Killing trajectories are timelike, namely outside the ergosphere in the Kerr spacetime).
The spatial metric $h(u)_{\alpha\beta}=P(u)_{\alpha\beta}$ corresponds to a Riemannian metric on the observer quotient space, and spacetime index raising and lowering on spatial tensor fields is accomplished equivalently with the spacetime metric or this spatial metric, and corresponds to the same activity on the quotient Riemannian manifold. 

The unit volume 3-form 
$\eta (u)_{\alpha\beta\gamma} 
= u^\delta \,\eta_{\delta\alpha\beta\gamma}$ 
may be used to define a spatial duality
operation for antisymmetric spatial tensor fields
\beq
\dualp{u} A_\alpha =\frac12 \eta(u)_{\alpha\beta\gamma}A^{\beta\gamma}\ ,
\qquad
\dualp{u} A_{\alpha\beta} =\eta(u)_{\alpha\beta\gamma}A^\gamma\ ,
\eeq
and a spatial cross product for a spatial vector ($X$) and a symmetric spatial tensor ($A$) and for two symmetric spatial tensor fields
($A, B$)
\beq
\fl\qquad
  [X \times_u A]^{\alpha\beta} = \eta (u)^{\,\gamma\delta (\alpha }
  X_{\gamma} A^{\beta)} {}_{\delta}\ ,\qquad
  [A\times_u B]_{\alpha} = \eta (u)_{\alpha\beta\gamma}
   A^{\beta} {}_{\delta}\, B^{\delta\gamma}\ ;
\eeq
the latter is just the dual of the matrix commutator.
In a similar way one can introduce a spatial inner product  for a spatial vector ($X$) and a symmetric spatial tensor ($A$) and for two spatial symmetric tensors ($A, \, B$)
\beq\fl\quad
[X\cdot A]_\alpha =h(u)_{\beta\mu}X^\beta A^\mu{}_\alpha 
= X_\mu A^\mu{}_\alpha\ , \
[A\cdot B]_{\alpha\beta} 
= A_{\alpha\mu}h(u)^{\mu\nu} B_{\nu\beta}
= A_{\alpha\mu}B^{\mu}{}_{\beta}\ .
\eeq
Spatial projection also leads to the following spatial differential operators \cite{mfg}
\begin{enumerate}
\item the spatial covariant derivative
\beq
\del (u)_\alpha X_\beta =P(u)^\mu_\alpha P(u)^\nu_\beta \del_\mu X_\nu \ ,
\eeq
which also allows the introduction of the
generalized divergence  and (symmetrized) curl operations for spatial symmetric tensor fields (see Appendix B) by
\beq\label{eq:scurl}
\fl\quad
  [\div (u) A]^\alpha 
= \del(u)^\beta A^\alpha {}_\beta \ ,\quad [\Scurl (u) A]^{\alpha\beta}
  =\eta(u)^{\gamma\delta\,(\,\alpha} \del(u)_\gamma A^{\beta\,)} {}_\delta \ ;
\eeq
\item the spatial Lie derivative along $u$ (temporal Lie derivative)
\beq
X_\alpha'=\Lie(u)_u X_\alpha = P(u)_\alpha^\beta \Lie_u X_\beta ;
\eeq
\item the spatial Fermi-Walker derivative along $u$ (temporal Fermi-Walker derivative)
\beq
\overcirc X_\alpha =\del(u)_u X_\alpha = P(u)_\alpha^\beta \del_u X_\beta \ .
\eeq
\end{enumerate}
These may be extended to multi-index objects in the usual way. 
For stationary spacetimes and observers following Killing trajectories, the spatial Lie derivative along $u$ gives zero when applied to stationary fields and so is a more convenient temporal derivative operator than the spatial Fermi-Walker derivative along $u$ used in stating tensor equations in the Ellis approach.

Both the vector-tensor cross product and the Scurl operator annihilate the pure trace part of the symmetric spatial tensor
\begin{eqnarray}
&& 
 X \times_u A = X \times_u A^{\rm(TF)}\ ,\quad
 \Scurl A = \Scurl A^{\rm(TF)}\ ,\nonumber\\
&&
A^{\rm(TF)}_{\alpha\beta} 
= A_{\alpha\beta} -{\textstyle\frac13} h(u)_{\alpha\beta} A^\gamma{}_\gamma
\ .
\end{eqnarray}
Furthermore one can show that
\beq\label{eq:align}
  X \times_u A^{\rm(TF)}=0 \rightarrow A^{\rm(TF)} = k [X\otimes X]^{\rm(TF)}\ ,
\eeq
so a vanishing cross product aligns $X$ and $A^{\rm(TF)}$ in the only way `alignment' makes sense in this context. 
(Since $X^\alpha X^\beta [X\times A^{\rm(TF)}]_{\alpha\beta} =0$, only four independent linear conditions are imposed on the five independent components of $A^{\rm(TF)}$, leaving this as the general solution.)

The vorticity vector field $\omega(u)^\alpha=\frac12\eta(u)^{\alpha\beta\gamma}\omega(u)_{\beta\gamma}$ is defined as the spatial dual of the vorticity 2-form (see appendix A). In index-free notation these will be denoted respectively by $\omega(u)$ and $\dualp{u}\omega(u)$. The exterior derivative of the 1-form $u_\alpha$ is then
\beq
du = - u\wedge a(u) + 2\dualp{u}\omega(u)\ ,
\eeq
letting the context determine the index position in index-free formulas. 
Any 2-form can be written in the form
\beq
  \sigma= u\wedge \sigma_{(E)} + \dualp{u}\sigma_{(M)}\ ,
\eeq
where the spatial vector fields $\sigma_{(E)},\sigma_{(M)}$ are called its electric and magnetic parts, 
so
the electric and magnetic parts of $du$ are then $-a(u)$ and $2\omega(u)$.

All of these operations on stationary spatial tensor fields correspond in a natural way to operations on tensor fields on the Riemannian 3-manifold quotient space of observer world lines involving its volume form and metric covariant derivative.
Excessive projection operator algebra can be conveniently avoided by working in a stationary frame $\{e_\top, e_a\}$ adapted
to the observer $e_\top=u$, i.e.\ $e_a \cdot u =0$.
The explicit dependence of fields on the observer $u$ can also be  suppressed to streamline the notation as will usually be done below and the following index conventions will be adopted: the index $\top$ indicates the temporal component, as in $X_\top = u \cdot X = - X^\top$, while Latin indices $a,b=1,2,3$ specify
spatial tensor components in an observer-adapted frame. As a convenience, purely spatial tensor equations will be stated with Latin indices below.

\section{The $1+3$ Maxwell-like equations for the Weyl tensor}

The Weyl tensor $C_{\alpha\beta\gamma\delta}$
is the part of the spacetime curvature which is not directly determined by the energy-momentum tensor.
Using its tracefree property, the once-contracted Bianchi identities for a generic $n$-dimensional manifold determine the divergence of the Weyl tensor \cite{Eisenhart}
\beq\eqalign{
   0 &= 3 R^{\alpha\beta}{}_{[\gamma\delta;\epsilon]} \delta^\gamma{}_\alpha
     = R^{\alpha\beta}{}_{\delta\epsilon;\alpha}
           + 2 R^\beta{}_{[\delta;\epsilon]} \cr
     &= C^{\alpha\beta}{}_{\delta\epsilon;\alpha}
           + (n-3)/(n-2)\, R^\beta{}_{\delta\epsilon} \ ,\cr
}\eeq
where
\beq
     R^\beta{}_{\delta\epsilon}
         = 2 ( R^\beta{}_{[\delta}
             -R/[2(n-1)] \delta^\beta{}_{[\delta} )_{;\epsilon]} \ ,
\eeq
thus expressing the divergence of the Weyl tensor for $n>3$
in terms of a
vector-valued 2-form ($R^\alpha{}_{\beta\delta}=-R^\alpha{}_{\delta\beta}$),
the Cotton tensor  \cite{Cotton},
which is the sign-reversed covariant exterior derivative
of the Ricci tensor plus a multiple of the scalar curvature times the
identity tensor seen as a vector-valued 1-form.

For $n=3$, the Weyl tensor vanishes identically and one can take the dual on the 2-form index pair of the Cotton tensor to obtain the Cotton-York tensor
$y$ \cite{York}, a divergence-free tracefree symmetric tensor which by the twice contracted Bianchi identities is equivalent to
\beq
 y_{ab}= -[\Scurl({\rm Ricci})]_{ab}\ ,
\eeq
where the $\Scurl$ operator is the obvious 3-dimensional operator corresponding to the spatial operator (\ref{eq:scurl}). Modulo a conformal rescaling, this tensor is conformally invariant (see Appendix C) and so vanishes for conformally flat geometries.

For $n=4$, half the Cotton tensor 
is a vector-valued current 2-form in the analogy with electromagnetism \cite{HE} 
(where $F^{\alpha\beta}{}_{;\beta} = 4\pi J^\alpha$)
\beq
\label{current}
 J^\alpha{}_{\beta\gamma} =C^{\alpha\delta}{}_{\beta\gamma;\delta} =
                           -\del_{[\beta} ( R^\alpha{}_{\gamma ]}
                           -\frac16 R \delta^\alpha{}_{\gamma ]} ) \ .
\eeq
Splitting these equations with respect to a generic
congruence $u$ leads to their $1+3$ Maxwell-like form  \cite{Ellis99}.
Note that either of these equivalent expressions defines $J^\alpha{}_{\beta\gamma}$, but one may impose the Einstein equations (in Ricci form) 
\beq
  R^\alpha{}_\beta 
 = \kappa(T^\alpha{}_\beta - {\textstyle\frac12} T^\gamma{}_\gamma\delta^\alpha{}_\beta)
\eeq
by using them to replace the Ricci tensor and scalar curvature terms in the second expression by the energy-momentum tensor, which vanishes in the vacuum spacetimes under consideration here, leading to source-free Maxwell-like equations.

First splitting the Weyl tensor  yields the two symmetric tracefree spatial fields
\beq
\fl\qquad
  E \,^{\alpha} {}_{\beta} 
=  C^{\alpha} {}_{\gamma\beta\delta}\, u^\gamma \, u^\delta \ ,\quad
  H \,^{\alpha} {}_{\beta} 
= - {}^\ast C^{\alpha} {}_{\gamma\beta\delta}\, u^\gamma \, u^\delta
= \frac{1}{2}\, \eta \, ^{\alpha} {}_{\gamma} {}^{\delta}\,
 C^{\gamma}{}_{\delta\beta\rho}\, u^\rho 
\ ,
\eeq
which are called its electric and magnetic parts respectively. These fields can be used to classify the gravitational
field by Petrov type  \cite{Kramer}.
Then splitting the Cotton tensor (\ref{current}), 
one finds in adapted frame component notation
\cite{Ellis99}
\beq\eqalign{
   J^\top{}_{a\top}
&=[\div E]_{a}+3 [\omega \cdot H]_a
+ [\theta \times H]_a 
\ ,\cr
   J^{*} {}^\top{}_{a\top}
&=-[\div H]_{a} + 3 [\omega \cdot E]_a
+[\theta \times E]_a
\ ,\cr
   J_{(\,ab\,)\,\top} 
&= [\Scurl H +2a\times H]_{ab} - E^{'}_{ab}
      -[\omega\times E]_{ab} 
\cr
      &- 2\Theta E_{ab} +5[\theta\cdot E]_{ab} -h_{ab}{\mathrm{Tr}}[\theta\cdot E]
\ ,\cr
   J^{*} {}_{(\,ab\,)\,\top} 
&= [\Scurl E +2a\times E]_{ab} + H^{'}_{ab}
       +[\omega\times H]_{ab} 
\cr
      &+ 2\Theta H_{ab} -5[\theta\cdot H]_{ab} 
  + h_{ab}{\mathrm{Tr}}[\theta\cdot H]
\ .\label{MaxLike1}
}\eeq
These are usually stated in terms of the spatial Fermi-Walker derivative along $u$ (open circle) rather than the spatial Lie derivative along $u$ (prime). To translate back one needs the relation
\beq
    \overcirc X_{ab}
         = X'_{ab} - 2 [\omega \times X]_{ab}
                   + 2 [\theta \cdot X]_{ab} \ .
\eeq

We now limit ourselves to the case of a stationary vacuum spacetime with the associated Killing vector $\xi$ and symmetry adapted coordinates $\{t,x^a\}$ with $\xi=\partial_t$ and employ
observers following the time coordinate lines (Killing trajectories).
These Killing observers have a vanishing expansion tensor and 
stationary fields $X$ then have a vanishing temporal Lie derivative $X^\prime=0$ with respect to them.
Where $\xi$ is timelike, the spacetime metric can be expressed in the form
\beq
ds^2=g_{\alpha\beta}dx^\alpha dx^\beta 
= -M^2 (dt-M_a dx^a)^2+\gamma_{ab}dx^a dx^b \ ,
\eeq
where $M=\sqrt{-g_{tt}}$ is the lapse factor, $M_a=-g_{ta}/g_{tt}$ is the shift 1-form and   $\gamma_{ab}=g_{ab}+M^2M_aM_b=h(m)_{ab}$ represents the spatial metric with respect to the chosen observers, whose 4-velocity
$u\equiv m=M^{-1}\partial_t$ is parallel to $\xi$. Their expansion is zero, while their acceleration and vorticity vector are
\beq
a=\vec\nabla \ln M\ ,\qquad \omega = {\textstyle \frac12} M\curl \vec M\ ,
\eeq
where $\vec M$ is an index-free way to distinguish the shift vector field from the lapse function.
The spatial covariant derivative symbol $\vec\nabla=\nabla(m)$ is adopted for the abbreviated notation in index-free formulas to distinguish it from the spacetime symbol $\nabla$; in Latin-indexed formulas $\vec\nabla$ will be denoted by $\nabla_a$. Note that $g=-a$ and $\vec H=2\omega$ are respectively the gravitoelectric and gravitomagnetic vector fields associated with these observers \cite{mfg}.

The Maxwell-like equations (\ref{MaxLike1}) with zero current in vacuum then reduce to
\begin{eqnarray}
\fl\qquad
&&
   \phantom{\,-\,\,} [\div E]_{a} + 3 [\omega \cdot H]_a=0\ ,\qquad
    [\Scurl H +2a\times H]_{ab} =[\omega\times E]_{ab} \ ,\nonumber\\
\fl\qquad
&&
    -[\div H]_{a} + 3 [\omega \cdot E]_a=0\ ,\qquad
    [\Scurl E +2a\times E]_{ab} =-[\omega\times H]_{ab} 
\ .\label{MaxLike2}
\end{eqnarray}
Defining the spatial complex fields
\beq
    Z=E-iH\ ,\qquad
    z=-a-i\omega = g-i\vec H/2 \ ,
\eeq
these equations assume the following compact form in index-free notation 
\beq
\div Z + 3i \omega \cdot Z =0\ ,\qquad
\Scurl Z +a\times Z = z\times Z \ ,
\eeq
but they still contain both the real and complex quantities.
From the identities ($n$ any number)
\beq\label{eq:scurlidentity}
\fl
\label{id1}
\Scurl [M^{n} X] = M^n[n\, a\times X+\Scurl X]
\ ,\quad
\div [M^{n} X] = M^n[n\, a\cdot X+\div X]
\eeq
valid for any (symmetric) 2-tensor field, these can be given their final form
\beq
\label{bianchi}
\div [M^{-3} Z] = 3M^{-3} z \cdot Z 
\ ,\quad
\Scurl [MZ] = M z\times Z
\ .\eeq
The complex combinations $z$ and $Z$ are associated with the $SO(3,C)$ representation of the Lorentz group exploited in the `self-dual formalism' \cite{Debeve,Taub,Kramer} closely related to the Newman-Penrose formalism.

These stationary sourcefree Maxwell-like equations are very similar to the stationary sourcefree Maxwell equations for a 2-form \cite{mfg} expressed in complex self-dual form
\begin{eqnarray}
\label{eq:max}
  && F=m\wedge \vec E + \dualp{m} B\ ,\ x = \vec E -iB\ ,\nonumber\\
  && \div(M^2 x) = -2 M^2 z\cdot x\ ,\
     \curl(M x)=0
\end{eqnarray}
except for the missing $z\times x$ term in the curl equation.
However, when $z$ and $Z$ are aligned, the correspondence is complete.

Let $R(h)_{ab}$ be the Ricci tensor of the spatial metric $h_{ab}$ with respect to the Killing observers, defined by the usual curvature formula involving spatial derivatives of $h_{ab}$. Then the vanishing expansion tensor of the Killing observers causes this tensor to agree with both the
symmetry-obeying spatial Ricci tensor $R_{\rm (sym)}{}_{ab}$ and the Lie spatial Ricci tensor introduced in  \cite{mfg} (section VII). 
The spatial part of the Einstein equations for a stationary vacuum spacetime then takes the form (see  \cite{Kramer}, Eqs.~16.25 and 16.26 or the final equation in (7.9) of \cite{mfg}, where $g_a=-a_a,H_a=2\omega_a$)
\beq
R(h){}^a{}_b=\nabla_b a^a +a^a a_b +2\omega^a \omega_b 
-2 \delta^a{}_b \omega^c \omega_c
\ .
\eeq
Since the Scurl annilates the pure trace part of spatial tensors, the Cotton-York tensor of the spatial metric $\gamma_{ab}$ is then
\beq\label{eq:cottonyork}
y=-\Scurl R(h)
= -\Scurl [\vec\nabla a +a\otimes a +2\omega\otimes  \omega]\ ,
\eeq
and by using the identities (see appendix C)
\begin{eqnarray}\label{eq:ScurlaR}
&&    \Scurl \vec\nabla a + a\times R(h) = 0\ , 
\nonumber\\
&&    \Scurl (a\otimes a)+ a\times \vec\nabla a
    =  0\ ,
\label{ident}
\end{eqnarray}
valid for any stationary vector field and arbitrary stationary gradient vector field respectively  (in particular for the acceleration vector,  a stationary gradient vector field)
and Eqs.~(\ref{id1}) one finds
\beq\label{eq:axRww}
y = 2a\times R(h) -2M^{-1}\Scurl [M\omega\otimes  \omega]\ .
\eeq

Finally let us examine the role played by the Cotton-York tensor in the Bianchi identities (\ref{bianchi}).
From
the explicit expressions (obtained directly from their definitions)
\beq\eqalign{
    E_{ab} &= \frac12 [\nabla_a a_b +a_aa_b 
          +R(h){}_{ab}-4\omega_a \omega_b]\TF \ ,\cr
    H_{ab} &=  [\nabla_a \omega_b +2a_a\omega_b]\TF \ ,\cr
}\eeq
and  by applying the $\Scurl$ operation and the identities (\ref{id1}) one finds

\beq
\label{scurleh}
\eqalign{
    \Scurl (ME) &= -\frac{M}{2} y -2\Scurl (M\omega\otimes \omega )
     ,\cr
    \Scurl (MH) &=  \Scurl [M^{-1}\vec\nabla (M^2 \omega )]\ .\cr
}\eeq
The $\Scurl$ Bianchi identities $M^{-1} \Scurl [MZ] = z\times Z$ then become
\beq
    W   -i  M^{-1}\Scurl [M^{-1}\vec\nabla (M^2 \omega )]
    = z\times Z\ .
\eeq
The real part $W$ of its left hand side can be expressed in a number of equivalent ways
\begin{eqnarray}\label{eq:ScurlZreal}
W 
&=& -{\textstyle\frac12} y -2 M^{-1}\Scurl (M\omega\otimes\omega )
= {\textstyle\frac12} [y-4a\times R(h)]\nonumber\\
&=& -{\textstyle\frac12} M^{-4}\Scurl (M^4 R(h))
= {\textstyle\frac12} [y+4\Scurl(\vec\nabla a)]
\ ,
\end{eqnarray}
where the second expression follows from the first using (\ref{eq:axRww}) and the third from the definition of the Cotton-York tensor $y$ and the first of the identities (\ref{eq:scurlidentity}) and the last one from the first of the identities (\ref{eq:ScurlaR}). 

\section{Papapetrou fields and the Simon-Mars tensor}

In a classic paper \cite{pap66} Papapetrou pointed out that a Killing vector field $\xi$ can be
interpreted as a vector potential satisfying a covariant Lorentz gauge and generating an electromagnetic field
\beq
\widetilde F_{\alpha\beta}
=[d \xi]_{\alpha\beta}
= \nabla_\alpha \xi_\beta - \nabla_\beta \xi_\alpha
= 2\nabla_\alpha \xi_\beta 
\eeq
solving Maxwell's equations with a current source which vanishes in vacuum spacetimes
\beq
\fl\qquad
 \xi_{(\alpha;\beta)}=0 
\rightarrow
 \xi^\alpha{}_{;\alpha}=0 
\rightarrow
 \xi_{\alpha;\beta}{}^{;\beta} = -\xi_\beta R^\beta{}_\alpha 
\rightarrow
\tilde F_{\alpha\beta}{}^{;\beta} = 2\xi_\beta R^\beta{}_\alpha =J_\alpha
\eeq
where the divergence condition is merely the trace of the Ricci identity for a Killing vector field.
Fayos and Sopuerta call $\tilde F$ a  `Papapetrou field' \cite{Fasop99}. Their remarkable result is that Papapetrou fields provide a
link between the Killing symmetries and the algebraic structure of the spacetime curvature which can
be found by studying the alignment of the principal null directions of the Papapetrou field
with those of the Riemann tensor itself. 
In the case of the Kerr spacetime, the
Papapetrou field generated by the timelike Killing vector field $\xi$ has its principal null directions
aligned with the two repeated principal
null directions of the spacetime \cite{Mars99,Fasop99} and this condition is equivalent to the vanishing of the Simon-Mars tensor,
a condition which in turn uniquely characterizes the spacetime itself in its symmetry class of stationary asymptotically flat spacetimes, as elegantly shown by Mars \cite{Mars99,Mars00}.

Consider first the case of a general stationary spacetime.
The Papapetrou field associated with $\xi =M m$ is $\tilde F = d(M m)$.
However, in the gravitoelectromagnetic language, it is more convenient to frame the discussion in terms of the observer 4-velocity $m$, so introduce the rescaled tensor field 
\beq
\fl\qquad
F=(2M)^{-1}\widetilde F ={\textstyle\frac12}M^{-1} d (Mm)
= {\textstyle\frac12} [dm - m\wedge a]= - m\wedge a +\dualp{m} \omega
\eeq
and its dual
\beq
\dual F=-m\wedge \omega -\dualp{m} a\ ,
\eeq
so that the complex 2-form
\beq
\mathcal{F}=F+i\rightdual F= m\wedge z +i\rightduals z\ ,
\label{Fantidual}
\eeq
is self-dual and its electric part is $z$.
Analogously, the self-dual part of the Weyl tensor
\beq
\mathcal{C}_{\alpha\beta\mu\nu}=C_{\alpha\beta\mu\nu}+i\rightdual C_{\alpha\beta\mu\nu}\ ,
\eeq
has the electric part
\beq
Z_{\mu\nu}=E_{\mu\nu}-iH_{\mu\nu}=m^\alpha m^\beta \mathcal{C}_{\alpha\mu\beta\nu}\ .
\eeq

The complex Papapetrou vector field $x=Mz$ corresponding to half the complex Papapetrou field $2-$form  $\mathcal{F}=2M F$ satisfies the source-free Maxwell's equations (\ref{eq:max}) in vacuum spacetimes
\beq
  \div(M^3z)=-2 M^3z\cdot z\ ,\
  \curl(M^2 z) = 0\ ,
\eeq
which are equivalent to the remaining time-time and time-space parts of the vacuum Einstein equations in Ricci form: $R_{\top\alpha}=0$. The curl equation implies that locally $M^2z$ is the gradient of a potential, namely the Ernst potential
\cite{Kramer}, while the divergence equation is the Ernst equation for this potential.

The spatial cross product of $z$ and $Z$ defines the (complex, symmetric and tracefree) Simon-Mars 2-tensor
\beq
\label{simon}
{}^{(2)}S= z\times Z\ .
\eeq
Apart from a scale factor this coincides with the spatial dual of the three index (vector-valued 2-form) spacetime Simon tensor recently discussed by Mars \cite{Mars99} and which corresponds as a spatial tensor on spacetime to the tensor on the observer quotient space
originally defined by Simon\cite{Simon}, the agreement holding only in the vacuum case as a result of the field equations.
Of course the Simon-Mars 2-tensor can be introduced in an arbitrary spacetime for any observer family.
The following correspondence translates the Simon and Mars notation into the gravitoelectromagnetic notation
\begin{eqnarray}
\fl\qquad
{\rm Mars:}\qquad & & 
  (\lambda, \omega_a, h_{ab}, \gamma_{ab}, \sigma_a, Y_{ab}, {\textstyle\frac12}\eta_a{}^{cd} S_{bcd} )
\\
\fl\qquad
\qquad\to & &
  (M^2, -2M^2\omega_a, h_{ab}, M^2 h_{ab}, -2 M^2 z_a, 2 M^2 Z_{ab}, 
4M^4\, {}^{(2)} S_{ab} ) \ .\nonumber  
\end{eqnarray}

Then as a consequence of the vacuum Bianchi identities (\ref{bianchi})
one finds the following representation of the Simon-Mars tensor in terms of the Cotton-York tensor $y$  of the observer quotient space metric, the lapse $M$ and the vorticity vector $\omega$
\beq\label{eq:2simon}
\fl\qquad
{}^{(2)}S= -\frac{y}{2}  -2M^{-1}\Scurl (M\omega\otimes \omega )
-i M^{-1}  \Scurl [M^{-1}\vec\nabla (M^2 \omega )] \ .
\eeq
The vanishing of the Simon-Mars tensor is equivalent to
the alignment of $z$ and $Z$ by (\ref{eq:align})
\beq
Z=k[z\otimes z]^{\rm(TF)} \ ,
\eeq
which forces the rescaled self-dual Weyl field $MZ$ to be Scurlfree,
like the corresponding Maxwell equation $\curl(M^2z)=0$ satisfied by
the complex Papapetrou vector field $Mz$.
%

The vanishing Scurl condition is in turn equivalent to the two real conditions
\beq\label{eq:zerosimon1}
 y= -4 M^{-1}\Scurl (M\omega\otimes \omega )\ ,\
 M^{-1} \Scurl [M^{-1}\vec\nabla (M^2 \omega )]=0\ .
\eeq
The first of these two real conditions is equivalent to the vanishing of any of the other three representations of $W$ given in (\ref{eq:ScurlZreal}).
For the Schwarzschild case where the vorticity is zero, this forces the Cotton-York tensor to vanish, making the quotient geometry, which coincides with the induced time hypersurface geometry in this case, conformally flat.

\section{The special case of a Kerr spacetime}

In Boyer-Lindquist coordinates the Kerr line element is \cite{mtw}
\beq\fl\quad
  ds^2 =  -\left(1-\frac{2\mathcal{M} r}{\rho^2}\right)dt^2
        - \frac{4 a\mathcal{M}r s^2}{\rho^2}dt d\phi
+ \frac{\rho^2}{\Delta}dr^2 +\rho^2d\theta^2
        + \frac{\Lambda s^2} {\rho^2}d\phi^2 \ ,
\eeq
where $s=\sin\theta$, $c=\cos\theta$ and
\beq
\fl\qquad
 \rho^2  =  r^2+a^2 c^2 \ ,\quad 
 \Delta  = r^2+a^2 -2\mathcal{M}r \ , \quad
 \Lambda = -\Delta a^2 s^2 +(r^2+a^2)^2\ .
\eeq
For the associated Killing observers with 4-velocity $m=M^{-1}\partial_t$, usually referred to as the static observers, the splitting quantities have the values
\beq
\fl\quad
(M, M_\phi ,\gamma_{rr},\gamma_{\theta\theta},\gamma_{\phi\phi})
=
\left(\left(1-2\frac{\mathcal{M}r}{\rho^2}\right)^{1/2},
-\frac{2\mathcal{M} ars^2}{\rho^2 -2\mathcal{M} r},
\frac{\rho^2}{\Delta},
\rho^2,
\frac{\Delta\rho^2s^2}{\rho^2 -2\mathcal{M} r}\right)
\ ,
\eeq
while
the acceleration vector $a$ and the vorticity vector $\omega$ 
have the following components
\beq
\fl
a= \frac{\mathcal{M} }{\rho^3 (\rho^2 -2\mathcal{M} r)} (\sqrt{\Delta}
     \epsilon, -2a^2rsc , 0)\ ,\
\omega=-\frac{\mathcal{M}}{\rho^3 (\rho^3 -2\mathcal{M} r)} 
       (\,2 \sqrt{\Delta} acr, as\epsilon, 0)
\eeq 
in the natural observer-adapted spatial orthonormal frame
\beq
\fl\qquad
e_{\hat r}=\gamma_{rr}^{-1/2} \partial_r
\ ,\ 
e_{\hat \theta}=\gamma_{\theta\theta}^{-1/2}\partial_\theta
\ ,\ 
e_{\hat \phi}=\gamma_{\phi\phi}^{-1/2} ( \partial_\phi+M_\phi \partial_t)
\ ,
\eeq
where the abbreviation $\epsilon = r^2-a^2c^2$ has been used.
Evaluating $z$ and $Z$ then leads to the components
\beq\label{eq:zZ}
\fl\quad
z = -\frac{\mathcal{M}(r-iac)^2}{\rho^3 (\rho^2 -2\mathcal{M} r)}
(\,{\sqrt{\Delta}}\,,\,-ias\,,\,0\,)
\ , \quad 
Z = -\frac{3}{\mathcal{M}} (r+iac) M^2 [z\otimes z]^{\TF}\ .
\eeq
Note that $z$ is proportial to a gradient
\beq
(\partial_{\hat r},\partial_{\hat\theta},\partial_{\hat\phi})[r+iac]
= -1/\rho \, (\sqrt{\Delta},-ias,0) \propto z \ .
\eeq

Since $Z$ is aligned with $z$ in the sense of (\ref{eq:align}) in the Kerr spacetime, the Simon-Mars tensor $z\times Z$ must vanish.
The vanishing of the Simon-Mars tensor is equivalent to the alignment of electric part of the self-dual Weyl tensor and the  complex
gravitoelectromagnetic vector field (i.e. the electric part of the self-dual Papapetrou field).
Mars \cite{Mars99} and Fayos and Sopuerta \cite{Fasop99}  state the same result
in terms of the alignment of the principal null directions of the self-dual Papapetrou field (\ref{Fantidual})
and the repeated principal null directions of the Weyl tensor.

Since $\vec\nabla(r+iac)\propto z$, then $\vec\nabla(r+iac) \times [z\otimes z]^{\rm(TF)} =0$, and because of the identity 
\beq
\Scurl [f A]=\vec\nabla f \times A + f \Scurl A \ ,
\eeq
when $z$ is substituted into $Z$ in (\ref{eq:zZ}) and $Z$ in turn into $\Scurl (M Z)=0$, one finds $\Scurl (M^3 z\otimes z)=0$.
The real and imaginary parts of 
this final equation lead to another pair of conditions satisfied in Kerr
\beq
\Scurl [M^3 (\omega\otimes \omega -a\otimes a)]=0 \ ,\
\Scurl [M^3 (a\otimes \omega +\omega\otimes a)]=0 \ .
\eeq

The vanishing of the Simon-Mars tensor forces the Cotton-York tensor to assume a nonzero value which can be represented in the various equivalent forms following from the vanishing of the expressions (\ref{eq:ScurlZreal}). Its only nonvanishing components are
\beq
\fl\qquad
   y^{r\phi}=  
\frac{6 a^2 \mathcal{M}^2}{[\,\rho^2\,(\,\rho^2 -2\mathcal{M} r\,)\,]^{\frac{5}{2}}}\,\Delta c \ ,\
   y^{\theta\phi}=  
-\frac{6a^2\mathcal{M}^2}{[\,\rho^2\,(\,\rho^2 -2\mathcal{M} r\,)\,]^{\frac{5}{2}}}\, (\,r-\mathcal{M} \,)\,s\ , 
\eeq
and those of the associated  vector one can introduce are
\begin{eqnarray}
\fl\qquad
y_{\hat \phi}
= y^{\hat r}{}_{\hat \phi}e_{\hat r}
 +y^{\hat \theta}{}_{\hat \phi} e_{\hat \theta} 
&=& \frac{6a^2\mathcal{M}^2s{\sqrt\Delta}}
     {[\,\rho\,(\,\rho^2 -2\mathcal{M} r\,)\,]^3}
[{\sqrt\Delta} c e_{\hat r}
-(r-\mathcal{M})s  e_{\hat \theta}] \nonumber\\
&=&\frac{3a^2\mathcal{M}^2}{[\,\rho^2\,(\,\rho^2 -2\mathcal{M} r\,)^3\,]} [\partial_{\hat\theta}({\Delta} s^2) e_{\hat r}-
\partial_{\hat  r}(\Delta s^2)
e_{\hat \theta}]\ ,
\end{eqnarray}
showing that this vector is proportional to a gradient.
Expanding the components of $y$ in the limit $a\to 0$, one finds
\beq
\fl
   (y^{r\phi},  y^{\theta\phi})
= \left(\frac{6c\mathcal{M}^2}{r^5(r^2-2\mathcal{M}r)^{3/2}} a^2 +O(a^4),
  -\frac{6s\mathcal{M}^2(r-\mathcal{M})}{r^5(r^2-2\mathcal{M}r)^{5/2}} a^2 +O(a^4)
\right)\ .
\eeq

This limiting behaviour is similar to that found by Price and Garat \cite{prigar} for the Cotton-York tensor of the induced metric on axisymmetric spatial slices slightly deformed from the usual time slices.
The nonvanishing Cotton-York tensor for the quotient geometry shows that for the usual Killing observers, the quotient geometry for Kerr is not conformally flat as it is in the Schwarzschild limit, similar to the situation for the induced geometry of the usual time slices.
Price and Garat showed that there do not exist any conformally flat axisymmetric time slices $t = f(a, r, \theta)$ with the limiting behavior $t=a F (r,\theta) + O(a^2)$ which forces the usual conformally flat slicing in the Schwarzschild limit. One could study whether a similar statement holds for the quotient space geometry of an axially symmetric observer congruence, but this is not trivial.

\section{Conclusions}

The Maxwell-like Bianchi identities for the Weyl curvature in a stationary spacetime have been re-expressed in complex self-dual form in terms of the now standard gravitoelectromagnetic variables associated with a Killing observer congruence. The role of the complex Simon-Mars tensor in these equations is then immediately apparent, and its vanishing has been expressed in terms of a
series of equivalent conditions involving the kinematical fields of the observer congruence and the Cotton-York tensor associated with the quotient space geometry. 
Furthermore, previous discussion concerning the alignment of the principal null directions of the Weyl tensor and the Papapetrou field associated with the timelike Killing congruence is equivalent to a simple proportionality between the electric part of the self-dual Weyl tensor and the trace-free self tensor product of the electric part of the self-dual Papapetrou field associated with the observer congruence, which is also an immediate consequence of the vanishing Simon-Mars tensor condition when it is expressed as a symmetric tracefree two index tensor.
In other words we have shown that `gravitoelectromagnetism at the curvature level' ($Z$) results from a (tracefree) tensor product of
`gravitoelectromagnetism at the connection level' ($z\otimes z$) only for the special case of the Kerr solution. 

As Mars has noted \cite{Mars99}, the nonvacuum Kerr-Newman charged black hole spacetime also has the same alignment of the principal null directions of the Weyl tensor, Papapetrou field and electromagnetic field. A computer algebra system calculation easily shows that the Simon-Mars tensor vanishes as well, although for nonvacuum spacetimes one must now distinguish between the Simon-Mars tensor as a spatial tensor and the original Simon spatial tensor, which differ by source terms. 
Further discussion of this case will be addressed in future work.

\appendix

\section{Definitions and conventions}

The sign conventions chosen for the spacetime and spatial volume element components
and the vorticity tensor and vector of a timelike congruence of curves
with unit tangent $u^\alpha$ make
comparison of different articles even by the same author problematic.
The Misner, Thorne and Wheeler conventions  \cite{mtw} for the volume elements finally adopted by
Ellis (see Eqn.~(6) of  \cite{Ellis99}), using hatted indices to indicate components
in an orthonormal frame, are
\beq
   \eta_{\hat0\hat1\hat2\hat3}=1 = \eta_{\hat1\hat2\hat3}\ ,
   \eta_{\alpha\beta\gamma}
     = u^{\delta} \eta_{\delta\alpha\beta\gamma} \ .
\eeq
However, the Ehlers-Ellis semi-colon convention for the vorticity tensor
consistently used in the past by the Ellis school (until recently, including  \cite{Maartens})
\beq
   u_{\alpha;\beta}
   = \omega^{\rm(E)}_{\alpha\beta} + \theta_{\alpha\beta} - a_\alpha u_\beta
\eeq
as opposed to the opposite-signed $\del$ convention (see Eqn.~(11) of  \cite{Ellis99})
\beq
  \del_\alpha u_\beta
   = \omega_{\alpha\beta} + \theta_{\alpha\beta} - u_\alpha a_\beta
\eeq
leads to the opposite sign compared with the Newtonian vorticity vector and tensor (which are half the curl of the 3-velocity field and its dual respectively), restored in  the $\del$ convention
used here
\beq
   \omega^\alpha = \frac12 \eta^{\alpha\beta\gamma}\omega_{\beta\gamma}
                 = \frac12[\curl u]^\alpha
  = -\omega^{\rm(E)}{}^\alpha
  = -\frac12 \eta^{\alpha\beta\gamma}\omega^{\rm(E)}_{\beta\gamma}
\ .
\eeq

The 4-volume element sign convention propagates to the signs of the
magnetic parts of a 2-form or curvature tensor since they involve
the dual operation.
The Thorne-Ellis98 conventions (see Eqn.~(5.40) of  \cite{Thorne}
and Eqns.~(20),(21) of  \cite{Ellis99}) in an observer-adapted orthonormal frame $\{e_{\hat0},e_{\hat a}\}$
are

\begin{eqnarray}
    && E_{\hat a} = F_{\hat a\hat0}
       \ ,\
       B_{\hat a} = -\dual F_{\hat a\hat0}
                  = \frac12 \eta_{\hat a\hat b\hat c} F^{\hat b\hat c}
              \ ,   \nonumber\\
    && E_{\hat a\hat b} = C_{\hat a\hat0\hat b\hat0}
         = \frac12 \eta_{\hat a\hat c\hat d} 
                  \dual C^{\hat c\hat d}{}_{\hat b\hat0}
         = -\frac14 \eta_{\hat a\hat c\hat d} \eta_{\hat b\hat f\hat g}
                         C^{\hat c\hat d\hat f\hat g} 
           \ ,\nonumber\\
    &&
       H_{\hat a\hat b} = -\dual C_{\hat a\hat0\hat b\hat0}
         = \frac12 \eta_{\hat a\hat c\hat d} 
                      C^{\hat c\hat d}{}_{\hat b\hat0}
\ ,
\end{eqnarray}
or equivalently
\begin{eqnarray}
&&
   E_\alpha = F_{\alpha\beta} u^\beta\ ,\
   B_\alpha = \frac12 \eta_\alpha{}^{\beta\gamma} F_{\beta\gamma}\ ,\nonumber\\
&&
   E_{\alpha\beta} = C_{\alpha\gamma\beta\delta} u^\gamma u^\delta\ ,\
   H_{\alpha\beta} = \frac12 \eta_\alpha{}^{\gamma\delta}
         C_{\gamma\delta}{}_{\alpha\epsilon} u^\epsilon\ .
\end{eqnarray}
The old Ehlers-Ellis 4-volume element convention found in
Maartens  \cite{Ellis73,Maartens} leads to the elimination
of the minus sign before the dual fields for the magnetic
parts of these fields. The present conventions differ from
Exercise 14.15 of  \cite{mtw} by changing the
sign of the electric part of Weyl.

For spatially projected derivatives, the following conventions are followed, spatially
projecting all indices after applying the corresponding spacetime derivative.
For example, for a vector field one has
\begin{eqnarray}
&&
  \bar\del_\beta X^\alpha
     = h^\alpha{}_\gamma h^\delta{}_\beta \del_\delta X^\gamma
  \ ,\nonumber\\
&&
  \del_{\rm(fw)} X^\alpha
     = h^\alpha{}_\beta \del_u X^\beta
     = \overcirc X{}^\alpha
     = \dot X{}^{\langle\alpha\rangle}
     = h^\alpha{}_\beta  \dot X{}^\beta
  \ ,\nonumber\\
&&
  \del_{\rm(lie)} X^\alpha
     = h^\alpha{}_\beta \pounds_u X^\beta
     = X'{}^\alpha \ .
\end{eqnarray}

The dot convention used by the Ellis school avoids the spatial projection
on all indices, preferring to distinguish the additional
application of spatial projection by angle-bracketing the indices (and removing the trace of symmetric 2-tensors). 
Including the spatial projection yields
the spatial Fermi-Walker derivative, which has the advantage
that the spatial metric / spatial projection tensors have zero derivative and so this
derivative commutes with index shifting of spatial fields
\begin{eqnarray}
&&
  \overcirc h_{\alpha\beta} = 0\ ,\
  \overcirc h{}^{\alpha\beta} = 0\ ,\
  \overcirc h{}^\alpha{}_\beta =0\ ,\nonumber\\
&&
   h'_{\alpha\beta} = 2\theta_{\alpha\beta}\ ,\
   h'{}^{\alpha\beta} = -2\theta^{\alpha\beta}\ ,\
   h'{}^\alpha{}_\beta = 0\ .
\end{eqnarray}
The spatial Lie derivative along $u$ does not commute with index shifting on spatial fields,
introducing extra expansion tensor terms, while the spatial Fermi-Walker derivative of a tracefree spatial
tensor is also tracefree (and spatial).

\section{The curl operators for 2-tensors in 3-dimensions}

For a vector-valued 1-form (mixed two-index tensor) in 3 dimensions, the spatial dual of the spatial covariant
exterior derivative leads to a new vector valued 1-form which might naturally
be called the spatial curl of the tensor
\beq
    [\curl T]^a{}_b
    =  \eta_b{}^{cd} \del_c T^a{}_d \ ,
\eeq
If the spatial tensor $A^a{}_b$ is antisymmetric, with
dual $[\dual A]^a$,
then this curl may be re-expressed as
\beq
    [\curl A]^a{}_b
    = \del(u)_b [\dual A]^a
         - \delta^a{}_b \del_c [\dual A]^c
\eeq
with trace $-2 \div \dual A$.
If one begins
with a symmetric such tensor $S^a{}_b$, the result is the dual of its covariant exterior derivative and is automatically
tracefree
\beq\eqalign{
    [\curl S]^a{}_b = [ \dual D S^a ]_b
    &=  \eta_b{}^{cd} \del_c S^a{}_d \ ,\cr
}\eeq
and for a pure trace such tensor $T^a{}_b = t\delta^a{}_b$,
the result is the dual of the exterior derivative of the scalar
coefficient,
modulo sign and index shifting
\beq
    [\curl T]^a{}_b
       = - \eta^a{}_b{}^c \del_c t \ .
\eeq

This operator can also be symmetrized, leading to
an operator that will be called the $\Scurl$ operator here, although the Ellis
school  \cite{Ellis99} justifiably calls it simply the curl in the context
restricted to symmetric rank 2 tensors
\beq
    [\Scurl T]^{ab}
    =  \eta^{cd(a} \del_c  T^{b)}{}_{d} \ ,
\eeq
It maps
the symmetric spatial vector-valued 1-forms
to tracefree symmetric such forms and annihilates their pure trace part
\beq\eqalign{
    [\Scurl S]^{ab}
    &=  \eta^{cd(a} \del_c S^{b)}{}_d =      [\Scurl S\TF]^{ab} \ .\cr
}\eeq
This allows the more streamlined expression (\ref{eq:cottonyork}) for the Cotton-York tensor as the sign-reversed Scurl of either the Ricci or Einstein tensors or their common tracefree part. The distinction between the curl and Scurl operators is important  when one considers the four possible spatial curvature tensors one can define for an arbitrary spacetime, since three of them do not have the usual symmetry properties of such tensors for observer congruences with nonzero vorticity and expansion tensors\cite{mfg}. In a similar way, using the symmetry-obeying curvature, one can introduce the Cotton-York tensor for any timelike congruence in any spacetime.

\section{Conformal transformations in 3 dimensions}

Let $U_a=\del_a U$, which satisfies
$$\del_a U_b = \del_a \del_b U = \del_b \del_a U = \del_b U_a$$
or $\eta^{abc} \del_b U_c=0$.
Then the standard conformal transformation laws for the metric, volume 3-form, the covariant derivative of a symmetric tensor $X$ and the Ricci curvature tensors in 3 dimensions are
\beq
  \tilde g_{ab} = e^{2U} g_{ab}\ ,\ 
  \tilde g{}^{ab} = e^{-2U} g^{ab}\ ,\ 
  \tilde g{}^{1/6} = e^U g^{1/6}\ ,
\eeq
\beq
  \tilde \eta_{abc} = e^{3U} \eta_{abc}\ ,\ 
  \tilde \eta{}^{abc} = e^{-3U} \eta{}^{abc}\ ,
\eeq
\beq
 \tilde\del_d X^b{}_c 
  = \del_d X^b{}_c + C^b{}_{de} X^e{}_c -C^e{}_{dc} X^b{}_e\ ,
\eeq
\beq
 C^c{}_{ab} = 2 \delta^c{}_{(a} U_{b)} - g_{ab} U^c\ ,\  
 C^c{}_{[ab]}=0\ ,
\eeq
\beq
  \tilde R_{ac} = R_{ac} - \del_a U_c +U_a U_c 
              + g_{ac}  ( - \del^e U_e - U^e U_e )\ ,
\eeq
\beq
    \tilde R^a{}_c = e^{-2U} \tilde R_{ac}\ ,\
  \tilde \del_d \tilde R^a{}_c = e^{-2U} (\del_d-2 U_d)  \tilde R_{ac} \ ,
\eeq
so that
\begin{eqnarray}
\fl
  g{}^{-5/6} \tilde g{}^{5/6} \tilde y^{ea}
&=&  g{}^{-5/6}  \tilde g{}^{5/6} \tilde \eta^{edc} 
       \tilde \del_d \tilde R{}^a{}_c
       =  \eta^{edc} (\del_d - 2 U_d) \tilde R{}^a{}_c \nonumber\\
\fl\quad        
&=& y^{ea} + [\Scurl(-\del \vec U + \vec U\otimes \vec U) 
+ \vec U \times ( \del \vec U - {\rm Ricci}) ]^{ea} \nonumber\\
\fl\quad        
&=& y^{ea} - [\Scurl(\del \vec U) + \vec U\times {\rm Ricci}]^{ea} + [\Scurl(\vec U\otimes \vec U)+ \vec U\times\del \vec U]^{ea}\ .
\end{eqnarray}
The second pair of  terms in  square  brackets is identically zero for a  gradient $U_a=\del_a  U$ just  by symmetry of the second  derivatives alone 
while the second pair is identically zero by a Ricci identity, leading
to the conformal invariance of the Cotton-York tensor-density $Y^{ea}=g{}^{5/6} y^{ea}$ (see exercise 21.22 of \cite{mtw}).

To derive the appropriate Ricci identity one must use the fact that in 3 dimensions, the double dual of the curvature tensor is the sign-reversed Einstein tensor. Starting from the Ricci identity for a vector field

\beq
  2\nabla_{[b} \nabla_{c]}  U^a = R^a{}_{dbc} U^d\ ,
\eeq
and symmetrizing the dual leads to
\begin{eqnarray}
&& \eta^{bc(a} \nabla_{[b} \nabla_{c]} U^{e)} 
 = {\textstyle\frac12} \eta^{bc(a}R^{e)}{}_{dbc} U^d
 = -{\textstyle\frac12} \eta^{bc(a} \eta^{e)}{}_{dg} G^{gm}\eta_{mbc} U^d
 \nonumber\\
&&
\qquad = - \eta^{(e}{}_{dg} U^d G^{a)g} 
 = - \eta^{(e}{}_{dg} U^d R^{a)g}\ ,
\end{eqnarray}
namely
\beq
 \Scurl \nabla \vec U = - \vec U \times {\rm Ricci}\ ,
\eeq
since the vector-tensor cross product is insensitive to the trace.


\end{document}